\documentclass[journal,twoside,web]{ieeecolor}
\usepackage{generic}
\usepackage{cite}
\usepackage{amsmath,amssymb,amsfonts}
\usepackage{algorithmic}
\usepackage{graphicx}
\usepackage{algorithm,algorithmic}
\usepackage{hyperref}
\hypersetup{hidelinks=true}
\usepackage{textcomp}
\usepackage{cite}
\usepackage{booktabs}
\usepackage{multirow}
\usepackage{float}

\def\BibTeX{{\rm B\kern-.05em{\sc i\kern-.025em b}\kern-.08em
    T\kern-.1667em\lower.7ex\hbox{E}\kern-.125emX}}
\begin{document}

\title{NeuroCLIP: A Multimodal Contrastive Learning Method for rTMS-treated Methamphetamine Addiction Analysis}

\author{Chengkai Wang$^{\dagger}$, Di Wu$^{\dagger}$, Yunsheng Liao, Wenyao Zheng, Ziyi Zeng, Xurong Gao, Hemmings Wu, Zhoule Zhu, Jie Yang, Lihua Zhong, Weiwei Cheng, Yun-Hsuan Chen$^{*}$, and Mohamad Sawan$^{*}$\, \IEEEmembership{Life Fellow, IEEE}
\thanks{Chengkai Wang, Di Wu, Yunsheng Liao, Wenyao Zheng, Ziyi Zeng, Xurong Gao, Jie Yang, Yun-Hsuan Chen, and Mohamad Sawan are with CenBRAIN Neurotech Center of Excellence, School of Engineering, Westlake University, Hangzhou 310030, China ($^{*}$Corresponding author, e-mail: chenyunxuan@westlake.edu.cn; sawan@westlake.edu.cn). (Chengkai Wang and Di Wu contributed equally.)}
\thanks{Ziyi Zeng is also with the School of Data Science, Xiamen University Malaysia, Selangor 43900, Malaysia.}
\thanks{Hemmings Wu and Zhoule Zhu are with Department of Neurosurgery, Second Affiliated Hospital, School of Medicine, Zhejiang University, Hangzhou 310009, China.}
\thanks{Lihua Zhong is with the Department of Education and Correction, Zhejiang Gongchen Compulsory Isolated Detoxification Center, Hangzhou 310011, China.
}
\thanks{Weiwei Cheng is with Zhejiang Liangzhu Compulsory Isolated Detoxification Center, Hangzhou 311115, China.}
}

\maketitle

\begin{abstract}
\label{sec:abstract}
Methamphetamine dependence poses a significant global health challenge, yet its assessment and the evaluation of treatments like repetitive transcranial magnetic stimulation (rTMS) frequently depend on subjective self-reports, which may introduce uncertainties. While objective neuroimaging modalities such as electroencephalography (EEG) and functional near-infrared spectroscopy (fNIRS) offer alternatives, their individual limitations and the reliance on conventional, often hand-crafted, feature extraction can compromise the reliability of derived biomarkers. To overcome these limitations, we propose NeuroCLIP, a novel deep learning framework integrating simultaneously recorded EEG and fNIRS data through a progressive learning strategy. This approach offers a robust and trustworthy biomarker for methamphetamine addiction. Validation experiments show that NeuroCLIP significantly improves \textcolor{black}{discriminative capabilities among the} methamphetamine-dependent individuals and healthy controls compared to models using either EEG or only fNIRS alone. Furthermore, the proposed framework facilitates objective, brain-based evaluation of rTMS treatment efficacy, demonstrating measurable shifts in neural patterns towards healthy control profiles after treatment. Critically, we establish the trustworthiness of the multimodal data-driven biomarker by showing its strong correlation with psychometrically validated craving scores. These findings suggest that biomarker derived from EEG-fNIRS data via NeuroCLIP offers enhanced robustness and reliability over single-modality approaches, providing a valuable tool for addiction neuroscience research and potentially improving clinical assessments.

\end{abstract}

\section{Introduction}
\label{sec:introduction}

\IEEEPARstart{M}{ethamphetamine} addiction represents a significant and growing public health concern globally. According to the United Nations Office on Drugs and Crime World Drug Report 2024, approximately 27\% of the estimated 292 million people who used drugs in the past year used amphetamine-type stimulants, with methamphetamine being the most prominent \cite{unodc2024}. The situation is particularly concentrated in regions like China, where recent national statistics indicated over 2 million registered drug users, more than half of whom use methamphetamine\cite{ye2022prevalence}. Despite its prevalence, the current treatment landscape for methamphetamine dependence remains limited \cite{siefried2020pharmacological}. Psychological and behavioral rehabilitation approaches are common \cite{moszczynska2021current} but show restricted long-term success; relapse rates exceed 60\% within the first year, and three-year remission rates are estimated to be only around 12\%~\cite{kamp2019effectiveness, mcketin2012evaluating}. Repetitive transcranial magnetic stimulation (rTMS), a non-invasive brain modulation technique, has emerged as a promising therapeutic strategy \cite{Paulus2020Neurobiology, zhang2019effects}. However, effectively evaluating its benefits presents distinct challenges \cite{su2020intermittent}.

A major challenge in both standard addiction assessment and the validation of rTMS treatment efficacy lies in the heavy reliance on self-report data~\cite{chang2022efficacy, khalili2021validity}. Clinical outcomes and addiction severity are frequently gauged using questionnaires and interviews, such as the Desire for Drug Questionnaire (DDQ) \cite{franken2002initial}. These methods are inherently subjective, susceptible to patient bias, mood fluctuations, and even intentional misrepresentation \cite{thomas2006addiction}. This creates a critical disparity between a patient's reported state and their underlying neural condition, potentially leading to inaccurate assessments, ineffective treatment adjustments, and adverse social consequences. While the pursuit of objective, brain-based biomarkers is crucial, traditional approaches often rely on hand-crafted heuristic biomarkers \cite{tsai2021intelligent}. These pre-defined features, derived from existing knowledge or simplified signal characteristics, may overlook complex, non-linear patterns indicative of addiction and can be less adaptable to individual variability or the nuances of treatment effects. Establishing robust, data-driven biomarkers is therefore necessary for reliable addiction analysis and accurate therapeutic outcome evaluation.

Neuroimaging modalities offer a path toward objective assessment, but single-modality approaches have inherent limitations \cite{rasero2021integrating}. Electroencephalography (EEG) captures neural electrical activity with excellent temporal resolution, which is essential for tracking the rapid neural correlates of cue reactivity \cite{zhou2024identification}. Yet, its limited spatial resolution presents significant challenges in pinpointing the precise subcortical networks, such as those integral to reward processing and craving, that are known to be dysregulated in addiction \cite{daly2019electroencephalography}. Functional near-infrared spectroscopy (fNIRS), on the other hand, provides better spatial localization of activity within the cerebral cortex by measuring hemodynamic responses \cite{pinti2020present}. This makes fNIRS valuable for examining regions like the prefrontal cortex, critically implicated in the executive dysfunctions and inhibitory control deficits characteristic of addiction \cite{carollo2022unfolding}. Nevertheless, fNIRS's reliance on the slower hemodynamic response means it inherently misses the initial, millisecond-scale neural firing crucial for understanding the immediate onset of cue-induced craving or rapid cognitive shifts \cite{pinti2020present}.

Lastly, even when a potential brain-based biomarker is identified, a critical challenge remains: its trustworthiness. Neurophysiological signals such as EEG and fNIRS exhibit high inter-subject variability~\cite{wu2024towards} and are sensitive to different pre-processing and experiment pipelines, which can make findings difficult to reproduce and inconsistent across studies~\cite{kinahan2024achieving, eastmond2022deep}. A truly robust biomarker should not only accurately classify patient groups but also demonstrate a meaningful alignment with established, external clinical measures of the disorder's severity \cite{fnih2016_evidentiary_framework}. This step, often overlooked in computational neuroscience studies, is essential for validating that a biomarker is not just statistically discriminative but also neuroscientifically and psychometrically  relevant \cite{parmigiani2023reliability}. 

\begin{figure}[htbp]
    \centering
    \includegraphics[width=\columnwidth]{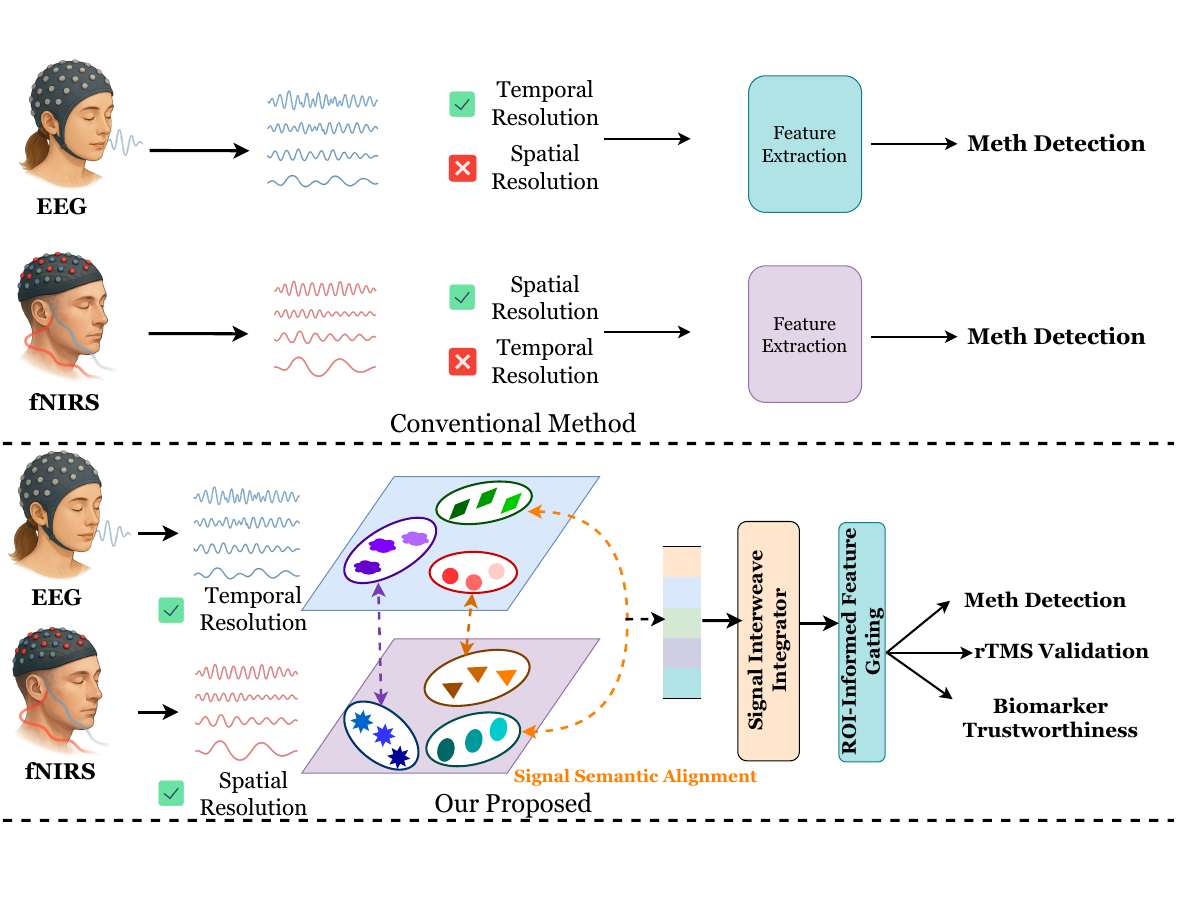}
    \caption{A conceptual comparison of multimodal analysis pipelines. Conventional methods analyze EEG and fNIRS data in isolation, forcing a trade-off between the high temporal resolution of EEG and the high spatial resolution of fNIRS. In contrast, our proposed framework, NeuroCLIP, employs a two-resolution process to synergistically integrate both signals. It first performs signal semantic alignment to project features from both modalities into a shared latent space. Subsequently, it integrates these aligned features to create a single, unified data-driven biomarker that leverages both high temporal and high spatial resolution simultaneously, enabling a more comprehensive and reliable addiction analysis.}
    \label{fig:teaser}
\end{figure}

To mitigate the previously discussed limitations, we propose NeuroCLIP, a novel end-to-end multimodal interpretable deep learning framework designed to integrate simultaneously acquired EEG and fNIRS signals for methamphetamine addiction analysis, in order to construct a more holistic and temporally precise understanding of the neural underpinnings of addiction and to extract robust data-driven biomarkers that capture the effects of therapeutic interventions. As conceptually illustrated in Figure \ref{fig:teaser}, NeuroCLIP overcomes the trade-offs of conventional methods by utilizing a progressive learning strategy to first learn aligned representations of the two modalities and then fuse them effectively to capture complementary information. Our model gains a deeper understanding of the neural semantics space through multiple stages, with the learned latent representations refined through various downstream tasks using specialized decoder heads. The framework aims to provide a more reliable analysis of addiction-related brain states, enable objective quantification of rTMS treatment efficacy beyond subjective reports and limited heuristic measures, and ultimately extract a robust and trustworthy multimodal data-driven biomarker for methamphetamine dependence from EEG-fNIRS data. \textcolor{black}{Furthermore, to verify the trustworthiness of the data-driven biomarker, we rigorously validate the derived biomarker against psychometrically established craving scores, ensuring its practical utility and relevance.}

In summary, the main contributions of this paper are:
{\color{black}
\begin{enumerate}
    \item We propose NeuroCLIP, a novel multimodal framework integrating EEG and fNIRS to identify brain-based biomarkers of methamphetamine addiction, enabling objective and quantitative evaluation of rTMS treatment beyond conventional self-reports or manually defined indicators.
    
    \item Implementation of interpretable analysis that quantitatively and reproducibly detects signal onset delays, addressing a critical gap in timing-sensitive neural analysis. To our knowledge, this is the first quantitative evaluation of signal onset delays in addiction studies.

    \item Achievement of an ROI-Informed Feature Gating unit that adaptively refines multimodal neural representations, enhancing relevant ROI-derived neural features while attenuating noise, sharpening focus on discriminative neurophysiological patterns.
    
    \item NeuroCLIP achieves state-of-the-art performance across several addiction-related predictive tasks. Importantly, the validity of NeuroCLIP's data-driven biomarkers is strongly confirmed by correlation analyses with psychometrically established addiction severity measures.
\end{enumerate}
}
The remainder of this paper is organized as follows. Section \ref{sec:related} reviews related work on methamphetamine addiction analysis and multimodal learning approaches. Section \ref{sec:task} describes the study participants, experimental protocol, and data collection procedures. Section \ref{sec:preliminary} introduces the concept of signal onset delay and presents our quantitative analysis of this phenomenon between EEG and fNIRS modalities. In Section \ref{sec:methods}, we provide a detailed description of our proposed NeuroCLIP framework while validating the trustworthiness of our proposed biomarker. Section \ref{sec:experiments} presents the experimental results, including addiction detection and rTMS efficacy evaluation. Finally, Section \ref{sec:discussion} discusses the implications of our findings and outlines directions for future research.

% In summary, the main contributions of this paper are:
% \begin{enumerate}
%     \item We propose NeuroCLIP, a novel multimodal analysis framework that integrates EEG and fNIRS signals to study methamphetamine addiction. This enables the discovery of brain-based biomarkers and provides a quantitative method to evaluate rTMS treatment effectiveness beyond conventional self-report measures and hand-crafted heuristic indicators.
    
%     \item An interpretable AI methodology is introduced that accurately detects signal onset delay through quantitative estimates, offering reproducible measurements critical for time-sensitive neural analysis. To the best of our knowledge, we are the first to quantitatively evaluate the signal onset delay for addiction studies.

%     \item An ROI-Informed Feature Gating unit employing sequential GELU and SwiGLU activation functions to adaptively refine fused multimodal representations. This mechanism enhances relevant ROI-derived neural features while attenuating noise, sharpening focus on discriminative neurophysiological patterns.
    
%     \item NeuroCLIP achieves state-of-the-art performance across multiple addiction-related analysis tasks. Furthermore, the trustworthiness of its discovered data-driven biomarker is validated through strong correlations with psychometrically established addiction severity measures.
% \end{enumerate}
\section{Related Work}
\label{sec:related}

\subsection{Methamphetamine Addiction Analysis}
Early research on methamphetamine (METH) use disorder (MUD) has relied on traditional signal analysis and machine learning applied to single brain modalities \cite{Chen2022Neuronal}. EEG microstate analysis partitions the EEG into quasi-stable topographical ``atoms of thought" states, and studies have reported that MUD patients show abnormal microstate dynamics compared to healthy controls \cite{Chen2020Disrupted, gao2025specific}. While microstates offer an interpretable window into large-scale brain network activity, these analyses are typically constrained to static EEG features and mixed-band data \cite{Koush2019Data-driven}, limiting their precision in isolating which frequency-specific or spatiotemporal components are affected by MUD~\cite{Zhao2023Deep}.

Researchers have applied classical statistical measures and machine learning classifiers to single-modality brain data \cite{Arora2018Comparison, yi2024nonparametric, yu2022outcome}. Hand-crafted EEG features, such as Power Spectral Density (PSD) or Event-Related Potentials (ERP) combined with algorithms like support vector machines or decision trees have shown moderate success in differentiating stimulant users from controls \cite{Ieracitano2019A, Abibullaev2021A}. However, these hand-crafted heuristic biomarkers are inherently limited by the pre-selected features; they may fail to capture more subtle or complex neural patterns indicative of addiction and often lack generalizability across different cognitive tasks or patient cohorts \cite{wu2024towards}. They tend to overfit to the specific conditions under which data were collected, and their interpretability hinges on the chosen features \cite{richards2019deep}. In contrast, data-driven approaches, particularly those using deep learning, can learn relevant features directly from the raw or minimally processed data, potentially uncovering more robust and nuanced biomarkers~\cite{wu2025towards}.

Beyond EEG, other sensing modalities have provided valuable but incomplete perspectives on methamphetamine addiction. Functional MRI and PET imaging have revealed neuroanatomical and metabolic alterations in MUD, yet these neuroimaging techniques are impractical for frequent monitoring \cite{sanjari2021effects}. Functional near-infrared spectroscopy (fNIRS) offers a middle ground – it noninvasively monitors cortical hemodynamics and has been used to differentiate drug cravings~\cite{chen2023challenges}. While fNIRS is low-cost and wearable, it samples at only ~10 Hz and captures delayed blood-flow responses, limiting its temporal acuity relative to EEG.

In summary, each single-modality approach comes with inherent trade-offs in sampling rate, spatial coverage, or invasiveness. Relying on any one modality often yields an incomplete picture of the addict's brain state, and many traditional analyses are offline or too simplified to guide in vivo neuromodulation.

\subsection{Multimodal Learning}
To overcome the shortcomings of single-modality analyses, researchers have turned to multimodal machine learning techniques that fuse information from different brain signals for enhanced biomarker discovery\cite{chen2023cinematic, lian2024driving}. Core challenges identified in multimodal learning include how to effectively represent each modality, synchronize or align the data, and fuse them without causing one to dominate or adding irreducible noise\cite{wang2024mmfusion}.

Within the neuroscience community, several works have proposed multimodal brain-signal fusion strategies, especially combining EEG with complementary modalities. A prominent line of research is hybrid EEG–fNIRS systems, since these can be worn together to monitor both fast electrical oscillations and slower hemodynamic trends from the cortex\cite{Liu2021A}. Such bimodal systems \cite{ali2023correlation} have shown that EEG and fNIRS can provide complementary information.

However, most early implementations merely concatenate features or use parallel classifiers for each modality, lacking a comprehensive systematic approach to properly fuse EEG and fNIRS data and exploit their complementary potential \cite{Shi2023Representing, mesgarani2012selective}. Besides, researchers often overlook the fundamental relationship between modalities, particularly the intrinsic signal onset delay between neural electrical activity and hemodynamic responses. While some studies acknowledge this delay, they primarily focus on compensating for it through temporal realignment rather than deeply understanding its neurophysiological basis \cite{chen2023cinematic}. 

Moreover, many existing multimodal systems are domain-specific or inflexible, as the fusion strategy tuned for one application (e.g. combining EEG and fNIRS for motor decoding) may not transfer to another (e.g. craving detection) \cite{wu2023transfer, andrews2024multimodal}.

The limitations of past studies highlight the necessity for a new multimodal learning paradigm tailored to the unique challenges of rTMS and MUD analysis – a gap that NeuroCLIP aims to fill.
\section{Participants and Task}
\label{sec:task}

\begin{figure}[tbp]
    \centering
    \includegraphics[width=\columnwidth]{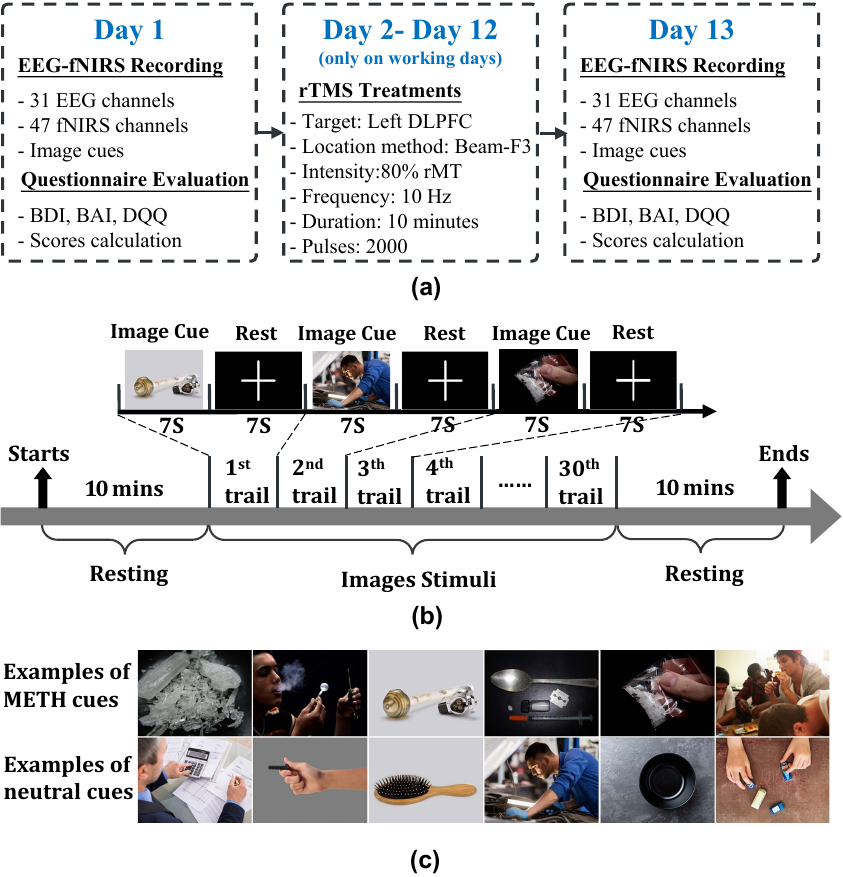}
    \caption{Overview of the experimental protocol. (a) Timeline for MUD participants, including baseline assessment (MBT), rTMS treatment period, and post-treatment assessment (MAT). (b) Structure of the cue-reactivity task with simultaneous EEG-fNIRS recording. (c) Examples of METH-related and neutral visual cues from the MOCD.}
    \label{fig:protocol}
\end{figure}

The experimental protocols, including participant groups, the overall timeline, and the specific tasks involved in data collection, are illustrated in Figure \ref{fig:protocol}. All procedures received approval from the ethical committees of Westlake University (ID: 20191023swan001) and The Second Affiliated Hospital Zhejiang University School of Medicine (ID: 2023\_0522) and were performed in strict adherence to the Declaration of Helsinki. Prior to inclusion in the study, written informed consent was obtained from all participants.

\subsection{Study participants}
Individuals diagnosed with methamphetamine use disorder (MUD) were initially recruited for this study from Zhejiang Gongchen Compulsory Isolated Detoxification Center. The diagnosis was established according to the Diagnostic and Statistical Manual of Mental Disorders, Fifth Edition (DSM-V) criteria. To be included, MUD participants needed to be between 18 and 50 years of age, have completed a minimum of 4 weeks of detoxification, express a clear intention to abstain from methamphetamine use, and demonstrate the capacity to understand and adhere to all study procedures. Individuals undergoing other concurrent treatments were excluded. From an initial cohort of 24 MUD recruits, 4 were excluded due to concurrent use of other substances, and an additional 3 were excluded because of poor signal quality in their recordings. The final MUD sample thus comprised 17 participants. The ages of these MUD participants ranged from 29 to 44 years, and all participants in this group were male. 

A healthy control (HC) group was concurrently recruited from Westlake university community via advertisements. The inclusion criteria for HC participants stipulated no history of methamphetamine use or any other substance use disorder, no current neurological or psychiatric conditions, and the ability to understand and comply with study protocols. Following the screening process, 17 HC participants were included for comparative analysis. The ages for the HC group ranged from 22 to 31 years and all of them are male.

\subsection{Task design}

For MUD participants, the study followed a longitudinal design spanning 13 days, as depicted in Figure \ref{fig:protocol}(a). The first day served as the pre-treatment baseline assessment. We refer to MUD participants at this stage as MBT (METH-addicted participants Before rTMS). On this day, MBT participants completed a set of clinical questionnaires: the Desire for Drug Questionnaire (DDQ), the Beck Depression Inventory (BDI) and the Beck Anxiety Inventory (BAI). Subsequent to these assessments, baseline EEG and fNIRS data were recorded simultaneously while participants performed a cue-reactivity task.

From Day 2 through Day 12, MUD participants underwent daily 10-minute sessions of rTMS treatment. This resulted in a total of 10 rTMS sessions, with scheduled breaks on weekends. On Day 13, designated as the post-treatment assessment day, MUD participants repeated the full battery of clinical questionnaires and underwent the identical EEG-fNIRS recording session, including the cue-reactivity task, that they had completed on Day 1. We refer to MUD participants at this stage as MAT (METH-addicted participants After rTMS).
HC participants attended a single experimental session. They were not required to complete any clinical questionnaires and did not receive any rTMS treatment. Their involvement was limited to the simultaneous EEG-fNIRS recording during the same cue-reactivity task performed by the MUD group.

\subsection{Data Collection}
The experimental paradigm employed for the simultaneous collection of EEG and fNIRS data is detailed in Figure \ref{fig:protocol}(b). Each recording session began with an initial 10-minute resting-state period, during which participants were instructed to sit calmly with their eyes closed while remaining awake. This initial rest phase was followed by the central cue-Reactivity task. In this task, participants were presented with a total of 30 visual stimuli, which included 15 images related to methamphetamine use and 15 neutral images. These visual cues were sourced from the Methamphetamine and Opioid Cue Database (MOCD) \cite{ekhtiari2020methamphetamine}, and their presentation was randomized. Figure \ref{fig:protocol}(c) provides representative examples of both the METH-related and neutral cues used. The structure of each trial involved displaying an image for 7 seconds. This image presentation was followed by a 7-second rest period, during which participants were instructed to maintain their focus on a central fixation cross displayed on the screen; this was intended to minimize head motion. The entire recording session concluded with a final 10-minute resting-state period, conducted under the same instructions as the initial resting phase.

\subsection{rTMS Treatment}
MUD participants received rTMS treatment once daily for 10 days, as indicated in the timeline in Figure \ref{fig:protocol}(a). The stimulation protocol utilized intermittent Theta Burst Stimulation (iTBS) and was adapted from procedures previously outlined by Liang et al. \cite{liang2018targeting}. Key parameters for the rTMS included a stimulation intensity set at 80\% of the individual’s active motor threshold (AMT). The iTBS protocol delivered 10 Hz bursts in a pattern of 5 seconds of stimulation followed by 10 seconds of rest. Each daily session had a total duration of 10 minutes, administering 2000 pulses. The rTMS was targeted at the left dorsolateral prefrontal cortex (DLPFC). To localize the left DLPFC, the Beam-F3 method was employed, with a round coil positioned 5 cm anterior to the scalp site corresponding to the motor threshold determined for the participant's contralateral hand. The rTMS session was initiated by clicking a start button and automatically stopped after the pre-programmed 10-minute duration.
\section{Preliminary}
\label{sec:preliminary}

\subsection{Signal Onset Delay}
Functional near-infrared spectroscopy (fNIRS) measures a hemodynamic response that is inherently delayed relative to the precise timing of underlying neuronal activity \cite{ferrari2012brief}. Characterizing this signal onset delay is important in addiction research, particularly when studying the brain's rapid responses to drug-related cues and the immediate effects of interventions, as it affects the accurate interpretation of neural event timing.

\subsection{Experimental Quantification}

To quantify modality-specific signal onset delays, we conduct an analysis focusing on the METH-addicted participants Before rTMS (MBT) group. Using the 7-second cue-presentation epochs from the task detailed in Section \ref{sec:task}.C, models employing the encoder architectures that are the backbones of NeuroCLIP (as detailed in Section \ref{sec:methods}, Figure \ref{fig:overview}) are trained independently on EEG data and fNIRS data to distinguish brain activity patterns associated with methamphetamine addiction, specifically by differentiating the MBT group from HC.

\textcolor{black}{To better understand when each modality begins contributing meaningful information toward addiction analysis, we analyze the temporal saliency of the derived feature. Specifically, we compute the gradients of the classifier's output probability for the correct class with respect to the activations in its final convolutional layer. This layer is chosen because it is known to encode rich, task-relevant semantic representations learned by the network~\cite{bashivan2019neural}.} For an input sample $s$ (an epoch of EEG or fNIRS data) and a target class $c$ (e.g., MBT), let $y_s^c$ be the model's output score for class $c$, and $A_{s,k}(t')$ be the activation of the $k$-th feature map (out of $N_K$) at temporal position $t'$ (out of $T'$) in the final convolutional layer. The influence of each feature map activation $A_{s,k}(t')$ on the class score $y_s^c$ is determined by its partial derivative: \begin{equation} G_{s,k}(t') = \frac{\partial y_s^c}{\partial A_{s,k}(t')} \end{equation} These gradients, $G_{s,k}(t')$, quantify the sensitivity of the class score to changes in the activations.

Next, channel-specific \textcolor{black}{saliency} weights, $\omega_{s,k}^c$, are calculated for each feature map $k$ by averaging these gradients $G_{s,k}(t')$ across all temporal positions $t'$: \begin{equation} \omega_{s,k}^c = \frac{1}{T'} \sum_{t'=1}^{T'} G_{s,k}(t') \end{equation} Each weight $\omega_{s,k}^c$ signifies the overall \textcolor{black}{saliency} of the $k$-th feature map in contributing to the model's score for class $c$ for sample $s$.

A temporal saliency map, $S_{s,k}^c(t')$, is then constructed for each sample $s$, channel $k$, and class $c$. This map results from a weighted linear combination of the forward activation maps $A_{s,k}(t')$, using the weights $\omega_{s,k}^c$, followed by ReLU to emphasize features positively influencing the class $c$ prediction: \begin{equation} S_{s,k}^c(t') = \text{ReLU}(\omega_{s,k}^c A_{s,k}(t')) \end{equation} The map $S_{s,k}^c(t')$ for each channel $k$ is a time series highlighting the most discriminative temporal segments in the feature space for classifying sample $s$ as class $c$.

To derive a temporal \textcolor{black}{saliency} profile from these channel-specific maps, we average across all channels, yielding a time series $I_s^c(t')$ that represents the \textcolor{black}{saliency} of each time point: \begin{equation} I_s^c(t') = \frac{1}{N_K} \sum_{k=1}^{N_K} S_{s,k}^c(t') \end{equation}

Finally, to achieve a robust, group-level understanding of temporal feature \textcolor{black}{saliency} for the MBT group, individual temporal \textcolor{black}{saliency} profiles $I_s^c(t')$ are averaged across all $N_{\text{MBT}}$ samples within the corresponding groups, producing an average temporal saliency profile, $\bar{I}^c(t')$: \begin{equation} \bar{I}^c(t') = \frac{1}{N_{\text{MBT}}} \sum_{s=1}^{N_{\text{MBT}}} I_s^c(t') \end{equation} 

The resulting temporal saliency profiles for MBT are depicted in Figure 2(a) for EEG and Figure 2(b) for fNIRS. To interpret these profiles, we consider a normalized saliency score exceeding 0.4 to be indicative of a substantial and consistent contribution from the modality's features to the network's decision-making process. As illustrated in Figure 2(a), the EEG-derived features rapidly gain saliency, with their score rising above the 0.4 level almost immediately after stimulus presentation. This aligns with expectations, as EEG directly captures voltage changes reflecting neural electrical activity in near real-time. Conversely, Figure 2(b) shows that the saliency of fNIRS-derived features reaches this informative level only at approximately 3 seconds post-stimulus. This observed delay in fNIRS reaching significant discriminative power reflects the natural latency of the hemodynamic response. Given the complementary nature of EEG (better temporal resolution) and fNIRS (better spatial resolution) in capturing distinct yet synergistic aspects of neural activity, a multimodal data-driven biomarker is crucial for accurately modeling brain responses related to addiction.

\begin{figure}[t]
\centering
\includegraphics[width=\columnwidth]{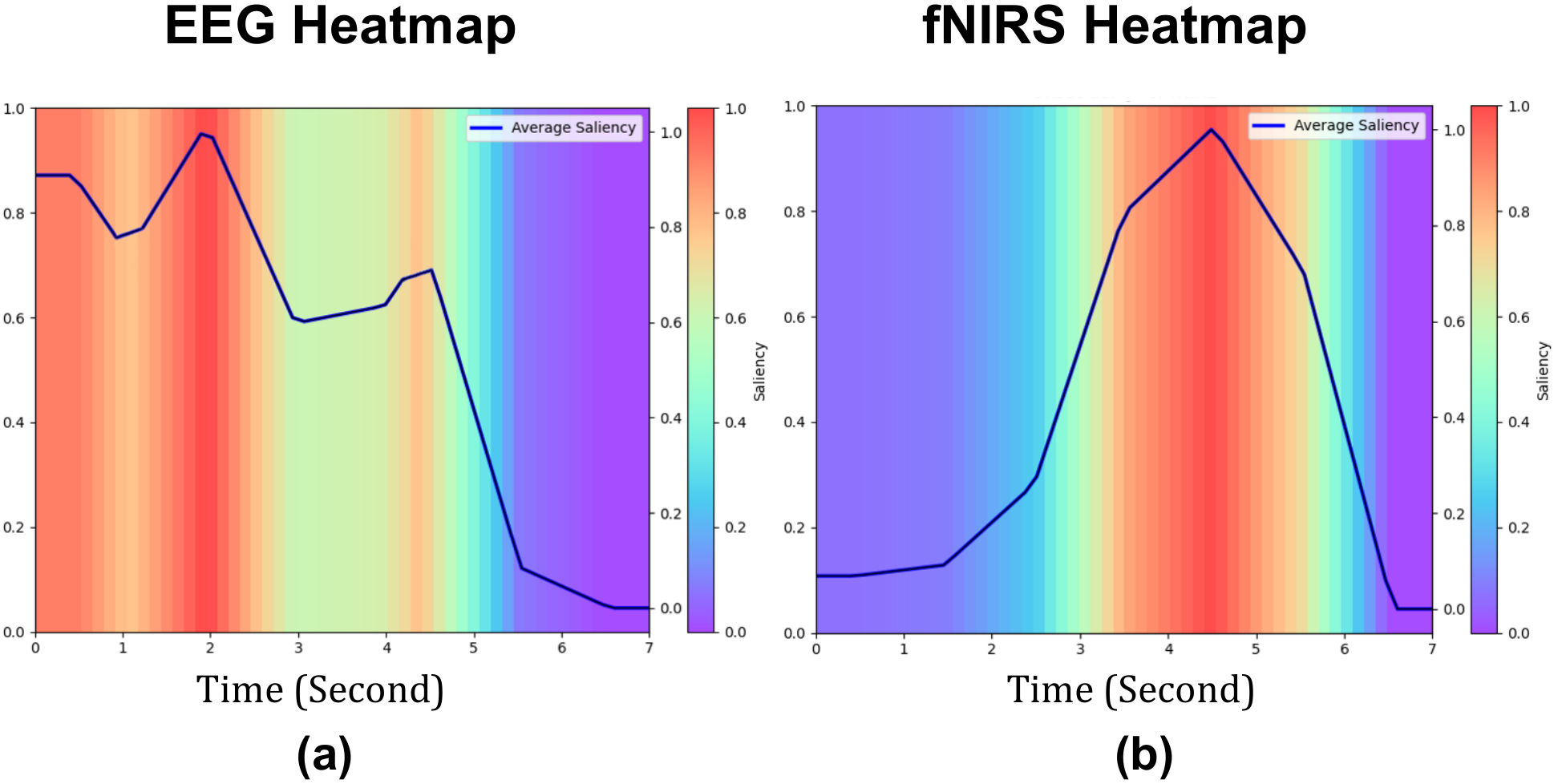}
\caption{Temporal saliency profiles showing signal onset delay between modalities. (a) EEG features gain discriminative saliency immediately upon stimulus presentation. (b) fNIRS features exhibit a 2.8-second delay before reaching a significant saliency level of 0.4, reflecting hemodynamic response latency.}
\end{figure}

\section{Methods}
\label{sec:methods}

\begin{figure*}[htbp]
    \centering
    \includegraphics[width=\textwidth]{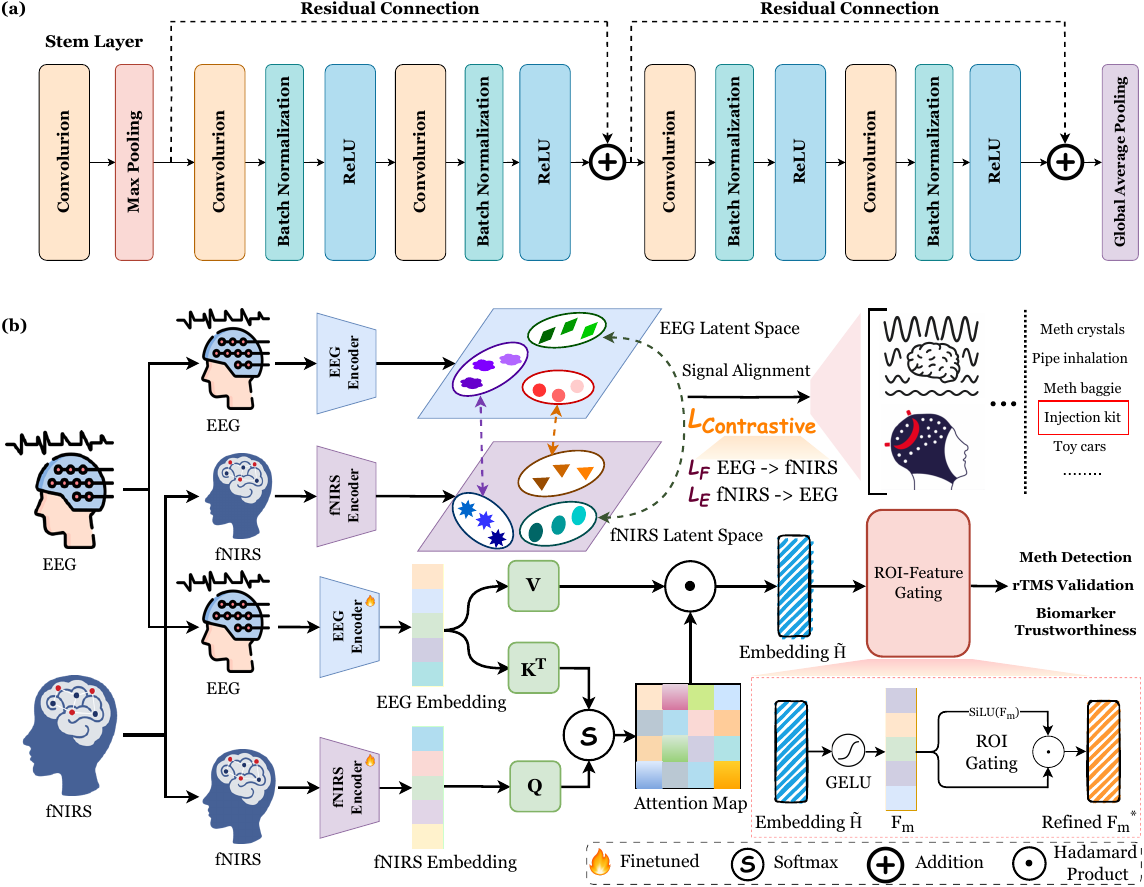}
    \caption{An overview of the NeuroCLIP framework. The model processes simultaneously recorded EEG and fNIRS data through three main stages. (a) The backbone for the modality-specific encoders is a ResNet-based Convolutional Neural Network. (b) In the Signal Contrastive Alignment Network, these encoders learn aligned representations in a shared latent space using a bidirectional contrastive loss. Subsequently, the Signal Interweave Integrator employs an attention-based mechanism to integrate the aligned features, capturing inter-modal dependencies. The fused representation is further refined by the ROI-Informed Feature Gating unit, which selectively enhances informative features. The final multimodal embedding is then used for various downstream decoding tasks.}
    \label{fig:overview}
\end{figure*}

We develop NeuroCLIP, a novel multimodal framework leveraging a progressive learning strategy. Progressive learning enables the model to first learn generalizable representations from each modality \textcolor{black}{with subsequent refinement} through task-specific cross-modal fusion and feature enhancement, aiming to generate meaningful multimodal embeddings tailored for addiction decoding. \textcolor{black}{The NeuroCLIP pipeline, illustrated in figure \ref{fig:overview}, comprises three main stages: the Signal Contrastive Alignment Network, the Signal Interweave Integrator, and the ROI-Informed Feature Gating unit. We detail each stage in the subsequent sections.}

\subsection{Signal Contrastive Alignment Network}

\textcolor{black}{The initial stage of NeuroCLIP aims to align the learned representations from multimodal signals.} This is achieved through a contrastive learning approach designed to project modality-specific features into a shared latent space where features from corresponding time segments are brought closer together, while features from non-corresponding segments are pushed apart. This alignment at a semantic level is crucial for effective downstream fusion.

\textcolor{black}{The encoder architecture for each modality can be flexibly selected according to the specific task requirements. For simplicity, in this work, we adopt a lightweight ResNet-based 1D Convolutional Neural Network (CNN) as the encoder backbone for both EEG and fNIRS modalities. The architecture is designed to particularly process the electrophysiological and hemodynamic time-series data for local temporal neural feature smoothing and denoising, preserving task-relevant temporal neural features.} Let $\mathbf{X}_e \in \mathbb{R}^{B \times D}$ and $\mathbf{X}_f \in \mathbb{R}^{B \times D}$ denote the L2-normalized feature embeddings extracted by these encoders from batches of EEG and fNIRS data, respectively, where $B$ is the batch size and $D$ denotes the embedding dimension (e.g., $D=32$). A pairwise similarity logits matrix, $\mathbf{S} \in \mathbb{R}^{B \times B}$, is computed between the EEG and fNIRS embeddings:
\begin{equation}
\mathbf{S} = \alpha\, \mathbf{X}_f\, \mathbf{X}_e^\top + \beta,
\end{equation}
where $\alpha = \exp(\tau)$ is a learnable scaling factor (with $\tau$ being a learnable parameter) with $\beta$ being a learnable scalar bias. Each element $\mathbf{S}_{ij}$ of $\mathbf{S}$ represents the similarity between the $i$-th fNIRS sample embedding ($\mathbf{x}_{f,i}$) and the $j$-th EEG sample embedding ($\mathbf{x}_{e,j}$): $\mathbf{S}_{ij} = \alpha\, (\mathbf{x}_{f,i})^\top \mathbf{x}_{e,j} + \beta$.

\textcolor{black}{The contrastive learning objective is based on the assumption that the representation derived from one modality for the $i^{th}$ sample in a batch should align closely with representations from other modalities for the same sample. The contrastive loss is therefore asymmetric, depending on the alignment direction. The loss of the alignment direction from fNIRS to EEG is defined as:}
\begin{equation}
\mathcal{L}_F = -\frac{1}{B} \sum_{i=1}^{B} \log \frac{\exp(s_{ii})}{\sum_{j=1}^{B} \exp(s_{ij})}
\end{equation}
Similarly, for the EEG-to-fNIRS direction, the loss is:
\begin{equation}
\mathcal{L}_E = -\frac{1}{B} \sum_{i=1}^{B} \log \frac{\exp(s_{ii})}{\sum_{j=1}^{B} \exp(s_{ji})}
\end{equation}
The overall contrastive loss, $\mathcal{L}_{\text{contrastive}}$, is the symmetric average of these two directional losses:
\begin{equation}
\mathcal{L}_{\text{contrastive}} = \frac{1}{2} (\mathcal{L}_F + \mathcal{L}_E)
\end{equation}
Minimizing this loss encourages the model to maximize the similarity for corresponding EEG-fNIRS pairs relative to non-corresponding pairs, thereby aligning the feature spaces of the two modalities.

\subsection{Signal Interweave Integrator}

Once aligned embeddings are obtained from the Signal Contrastive Alignment Network, the Signal Interweave Integrator employs a cross-attention mechanism to learn dynamic relationships and fuse information between the two modalities. This allows the model to adaptively weight the importance of features from each modality and capture long-range temporal dependencies crucial for modeling complex brain responses in addiction.

Given the aligned EEG embeddings $\mathbf{X}_e^{\prime}$ and fNIRS embeddings $\mathbf{X}_f^{\prime}$, we define the Query matrix $\mathbf{Q} = \mathbf{X}_e^{\prime}$, and the Key and Value matrices $\mathbf{K} = \mathbf{X}_f^{\prime}$ and $\mathbf{V} = \mathbf{X}_f^{\prime}$. This configuration allows EEG features to query relevant information from fNIRS features. A multi-head attention mechanism with $h$ heads (e.g., $h=4$) is used. For each head $i \in \{1, \dots, h\}$, learnable projection matrices $\mathbf{W}_i^Q, \mathbf{W}_i^K, \mathbf{W}_i^V \in \mathbb{R}^{D \times d_k}$ (where $d_k = D/h$ is the per-head dimension) are used to project $\mathbf{Q}$, $\mathbf{K}$, and $\mathbf{V}$ into subspaces:
\begin{align}
\mathbf{Q}_i &= \mathbf{X}_e^{\prime} \mathbf{W}_i^Q \\
\mathbf{K}_i &= \mathbf{X}_f^{\prime} \mathbf{W}_i^K \\
\mathbf{V}_i &= \mathbf{X}_f^{\prime} \mathbf{W}_i^V
\end{align}
The attention output for each head is computed as:
\begin{equation}
\text{head}_i = \text{softmax}\left(\frac{\mathbf{Q}_i \mathbf{K}_i^\top}{\sqrt{d_k}}\right)\mathbf{V}_i
\end{equation}
The outputs of all heads are then concatenated and passed through a final linear projection layer with weight matrix $\mathbf{W}_O \in \mathbb{R}^{D \times D}$ to produce the fused multimodal representation $\widetilde{\mathbf{H}} \in \mathbb{R}^{N \times D}$:
\begin{align}
\mathbf{H}_{\text{concat}} &= \text{Concat}(\text{head}_1, \dots, \text{head}_h) \\
\widetilde{\mathbf{H}} &= \mathbf{H}_{\text{concat}} \mathbf{W}_O
\end{align}
This fused representation $\widetilde{\mathbf{H}}$ encapsulates integrated information from both EEG and fNIRS.

\subsection{ROI-Informed Feature Gating}

To further refine the fused multimodal representation $\widetilde{\mathbf{H}}$ obtained from the Signal Interweave Integrator, we introduce an ROI-Informed Feature Gating Unit. This lightweight module aims to enhance the nonlinear expressiveness of the fused features, improve regularization, and enable selectivity over the learned feature dimensions. The goal is to emphasize features that are most informative for the task, which we hypothesize are strongly influenced by neural activity originating from high-contributing Regions of Interest (ROIs), while suppressing features dominated by noise or less relevant neurophysiological sources. 

The module operates sequentially. First, the Gaussian Error Linear Unit (GELU) activation is applied to the fused representation $\widetilde{\mathbf{H}}$:
\begin{equation}
\mathbf{F}_{\text{m}} = \text{GELU}(\widetilde{\mathbf{H}})
\end{equation}
Unlike activation functions like ReLU that perform hard thresholding, GELU introduces a smooth, probabilistic gating. This characteristic allows it to retain weak but potentially informative gradients in low-activation regions, which is beneficial for preserving subtle cross-modal dependencies that are crucial for capturing complex ROI interactions.

However, while GELU enhances the representation, the output $\mathbf{F}_{\text{m}}$ may still contain features that are noisy or less critical for addiction-related task, calling for an automatic feature selection mechanism. To address this, we subsequently apply SwiGLU (Swish-Gated Linear Unit) to \(\mathbf{F}_m\) as follows:
\begin{equation}
    \mathbf{F}_m^* = \text{SwiGLU}(\mathbf{F}_m) =
    \underbrace{\text{SiLU}(\mathbf{F}_m \mathbf{W})}_{\text{Gating branch}} \odot 
    \underbrace{(\mathbf{F}_m \mathbf{V})}_{\text{Linear branch}}
\end{equation}
Here, \(\mathbf{W},\mathbf{V}\!\in\!\mathbb{R}^{D\times D}\) are learnable weight matrices and \(\odot\) denotes element‑wise multiplication. The term $\mathbf{G} = \text{SiLU}(\mathbf{F}_m \mathbf{W})$ acts as the ``gating branch", and $\mathbf{L} = (\mathbf{F}_m \mathbf{V})$ serves as the ``linear branch" providing the content for each of the $D$ feature dimensions. The SiLU activation function ($\text{SiLU}(x) = x \cdot \sigma(x)$, where $\sigma(x)$ is the sigmoid function) in the gating branch produces values $\mathbf{G}_j$ for each feature dimension $j$ that can range from approximately 0 to values greater than 1. When a particular feature dimension's gate value $\mathbf{G}_j$ is close to 0, it effectively ``closes the gate" for that feature, suppressing its corresponding content $\mathbf{L}_j$ from the linear branch. Conversely, when $\mathbf{G}_j$ is significantly positive, the gate ``opens", allowing the feature content $\mathbf{L}_j$ to pass through, potentially amplified if $\mathbf{G}_j > 1$. This dynamic, data-driven gating allows the network to learn to assign higher relevance to informative features that represent high-contributing ROIs and lower relevance to features that reflect noise or less critical ROI activity. The outcome $\mathbf{F}_m^*$ is thus a refined representation where selection over feature dimensions has occurred, leading to improved gradient stability and a more focused feature set.

Together, the GELU and SwiGLU layers form a shallow feedforward transformation operating on the output of the cross-attention stage. This combination effectively enhances the model's nonlinear projection capabilities and improves generalization through adaptive feature selection. By refining the fused representation in this manner, the model can better leverage information that distinguishes addiction-related brain states, with the selection process guided by the learned importance of features reflecting differential ROI contributions. The enhanced representation $\mathbf{F}_m^*$ is then passed to a final MLP decoder for the downstream task.

\subsection{Biomarker Trustworthiness Validation}
 Establishing the trustworthiness of a brain-based biomarker is a critical step beyond demonstrating its ability to solve neural tasks. A robust biomarker should also show a meaningful quantitative alignment with established clinical indicators of a disorder's severity, ensuring it is not just statistically discriminative but also neuroscientifically relevant. To address this, we design an experiment to test if the multimodal biomarker extracted by NeuroCLIP from EEG-fNIRS data correlates with externally validated, standardized craving scores. For this purpose, we again leverage the Methamphetamine and Opioid Cue Database (MOCD), which provides images annotated with psychometrically verified craving scores. By testing our biomarker against this external standard, we can verify its trustworthiness beyond simple classification accuracy.

\begin{table}[tbp]
\centering
\caption{Ground Truth Categorization of MOCD Stimuli Based on Psychometrically Verified Craving Scores.}
\label{tab:craving_score_levels}
\small
\resizebox{\columnwidth}{!}{
\begin{tabular}{cccc}
\toprule
    Level & Image ID & Category & Craving Score \\
\midrule
High   & Meth13  & Meth                  & 94.14 \\
High   & Meth11  & Meth                  & 94.00 \\
High   & Meth23  & Meth Hand             & 93.93 \\
High   & Meth43  & Meth Instrument       & 93.79 \\
High   & Meth77  & Meth Instrument and hand & 93.64 \\
\midrule
Medium & Meth98  & Meth Injection and hand & 93.21 \\
Medium & Meth62  & Meth Instrument       & 93.11 \\
Medium & Meth48  & Meth Instrument       & 92.54 \\
Medium & Meth101 & Meth Face and activities & 92.50 \\
Medium & Meth68 & Meth Instrument       & 92.46 \\
\midrule
Low    & Meth46 & Meth Instrument       & 92.40 \\
Low    & Meth94 & Meth Injection and hand & 92.21 \\
Low    & Meth73 & Meth Instrument and hand & 92.21 \\
Low    & Meth69 & Meth Instrument and hand & 92.18 \\
Low    & Meth01 & Meth                  & 92.07 \\
\bottomrule
\end{tabular}
}
\end{table}

\begin{table}[!htbp]
\centering
\caption{Quantitative Alignment of the NeuroCLIP Biomarker with Standardized Craving Levels.}
\label{tab:classification_performance}
\small
\resizebox{\columnwidth}{!}{
\begin{tabular}{ccccc}
\toprule
Modality & Accuracy (\%) & Precision (\%) & F1 (\%) & Recall (\%) \\
\midrule
EEG-fNIRS & 89.74 & 91.03 & 95.65 & 92.31 \\
\bottomrule
\end{tabular}
}
\end{table}

While absolute craving scores naturally vary between individuals and study cohorts, the MOCD provides a standardized set of stimuli with a well-defined rank-ordering of craving-inducing potential. Our objective is therefore not to predict the exact numerical scores from the MOCD study, but to test a more fundamental hypothesis: can the neurophysiological biomarker extracted by NeuroCLIP from our participant cohort correctly classify these standard stimuli into their established low, medium, and high craving-potential categories? Success in this task would provide strong evidence that our biomarker is sensitive to the same underlying neural processes that these clinical scores represent.

For the experiment, we use the 15 METH-related images from the MOCD presented to our participants. Based on their associated craving scores from the MOCD study, these images are divided into three balanced categories: High Craving (top 5 scores), Medium Craving (middle 5 scores), and Low Craving (bottom 5 scores), as detailed in Table \ref{tab:craving_score_levels}
. NeuroCLIP is then trained on the corresponding 7-second EEG-fNIRS epochs from our participants to perform a three-way classification task predicting these labels.

The classification performance of NeuroCLIP on this task is presented in Table \ref{tab:classification_performance}. The model achieves high performance across all metrics, with an accuracy of 89.74\% and an F1-score of 95.65\%. This result is highly significant, as it indicates that the multimodal data driven biomarker extracted by NeuroCLIP is not only discriminative in a general sense but is also sufficiently sensitive enough to capture the nuanced differences in neural responses to stimuli with varying degrees of craving potential. This strong quantitative alignment with an external, psychometrically verified standard provides compelling evidence for the trustworthiness and clinical relevance of our discovered biomarker.

\section{Experiments} 
\label{sec:experiments}
\begin{table*}[htbp]
  \centering
  \caption{Single-modality performance on methamphetamine addiction detection compared to state-of-the-art methods.}
  \label{tab:sota_comparison_single_modality}
  \resizebox{\textwidth}{!}{%
  \color{black}\begin{tabular}{ccccccc}
    \toprule
    \multirow{2}{*}{Modality} & \multirow{2}{*}{Metrics} & SPaRCNet~\cite{jing2023development} & FFCL~\cite{li2022motor} & ST-Transformer~\cite{song2021transformer} & TS-TCC~\cite{ijcai2021tstcc} & This work \\
    & & (CNN) & (CNN+LSTM) & (Transformer) & (CNN) & (CNN) \\
    \midrule
    \multirow{2}{*}{EEG} & Accuracy (\%)      & 87.48$\pm$2.53        & 86.15$\pm$3.11         & 89.03$\pm$2.15            & 88.21$\pm$2.24        & \textbf{93.50}$\pm$\textbf{0.73} \\
                        & Sensitivity (\%)   & 86.12$\pm$2.78        & 84.67$\pm$3.24         & 87.55$\pm$2.29            & 86.49$\pm$2.51        & \textbf{91.05}$\pm$\textbf{1.71} \\
    \midrule
    \multirow{2}{*}{fNIRS} & Accuracy (\%)    & 83.17$\pm$3.05        & 82.33$\pm$3.48         & 85.12$\pm$2.76            & 84.68$\pm$2.65        & \textbf{90.10}$\pm$\textbf{1.47} \\
                           & Sensitivity (\%) & 81.62$\pm$3.18        & 80.19$\pm$3.75         & 84.08$\pm$3.01            & 83.24$\pm$2.93        & \textbf{92.59}$\pm$\textbf{1.86} \\
    \bottomrule
  \end{tabular}
  }
\end{table*}
\subsection{Data Pre-processing}
Data acquisition is performed using a NIRSport2 system (NIRx Medical Technologies LLC, USA) for fNIRS, employing dual-wavelength sources (760/850 nm). EEG data are recorded using 31 actiCAP slim active electrodes (Brain Products GmbH, Germany) connected to an actiCHamp Plus amplifier. Both modalities are configured according to the international 10–20 system for electrode and optode placement.

Raw EEG signals, initially recorded at a sampling rate of 1000 Hz, are first processed with a 4-45 Hz bandpass filter to eliminate low-frequency drift and high-frequency noise while also preventing aliasing. The data are then downsampled to 250 Hz to reduce dimensionality and improve computational efficiency. The ICLabel algorithm automatically identifies and removes components associated with eye movements, muscle activity and other noise sources with a rigorous manual inspection. Following these cleaning steps, the continuous EEG data are segmented into 7-second epochs corresponding to the visual cue presentation periods of the experimental task. Finally, each channel within each epoch is z-score normalized to avoid overfitting.

For fNIRS data, raw optical density measurements are converted into changes in oxyhemoglobin (HbO) using the Modified Beer-Lambert Law. The resulting data are bandpass filtered between 0.01 Hz and 0.2 Hz to preserve slow hemodynamic fluctuations related to neural activity while removing physiological artifacts such as respiratory and cardiac signals, as well as instrumental drifts. The fNIRS channels are mapped to eight predefined regions of interest (ROIs): Dorsolateral Prefrontal Cortex (DLPFC), Frontal Eye Fields (FEF), Motor Cortex, Left and Right Broca's Areas, Left and Right Temporal Cortices, and Visual Cortex. Data from the Visual Cortex ROI are excluded from subsequent model input due to concerns about signal quality and potential activation by general visual stimuli not specific to addiction. Any low-quality fNIRS channels are imputed using ROI-level averaging to maintain spatial consistency. Similar to the EEG data, the pre-processed fNIRS HbO signals are segmented into 7-second epochs aligned with the cue presentation, and each channel within each epoch undergoes z-score normalization.

\subsection{Methamphetamine Addiction Detection}

This experiment evaluates the fundamental capability of NeuroCLIP to distinguish brain activity patterns in methamphetamine-dependent individuals from those in healthy controls, thereby assessing the quality of biomarkers extracted from the multimodal data. We structure our evaluation in two parts. First, to establish the effectiveness of our chosen encoder backbone, we benchmark the performance of our 1D ResNet-CNN backbone against several state-of-the-art (SOTA) deep learning models on the single-modality addiction detection task. Second, we demonstrate the benefit of our proposed multimodal integration by comparing the full NeuroCLIP framework against single-modality baselines.

To validate our single-modality backbone, we compare its performance against representative SOTA models designed for neurophysiological time-series analysis. These benchmarks include: SPaRCNet~\cite{jing2023development}, a robust CNN architecture designed for EEG pattern classification; FFCL~\cite{li2022motor}, which combines a CNN with an LSTM to capture both spatial and temporal features; ST-Transformer~\cite{song2021transformer}, a Transformer-based model capable of learning long-range dependencies in EEG data; and TS-TCC~\cite{ijcai2021tstcc}, a contrastive learning framework for time-series representation. As shown in Table \ref{tab:sota_comparison_single_modality}, our 1D ResNet-CNN backbone consistently outperforms these established architectures on both EEG and fNIRS data for this specific task. This result confirms that our selected backbone is not merely a simplistic choice but a highly effective and well-suited feature extractor for addiction-related neural patterns, providing a strong foundation for subsequent multimodal integration.

Having established the strength of our backbone, we then quantify the benefit of multimodal integration by comparing the performance of the full NeuroCLIP framework against two single-modality baselines. The EEG-only and fNIRS-only baseline models employ the same respective 1D ResNet-CNN encoder architectures as NeuroCLIP. However, for these baselines, the extracted features are passed directly to a final MLP decoder, bypassing the Signal Contrastive Alignment, Signal Interweave Integrator, and ROI-Informed Feature Gating stages that constitute the full multimodal architecture for enhanced biomarker extraction.

\begin{table}[t]
\caption{Methamphetamine Addiction Detection Performance of NeuroCLIP versus Single-Modality Baselines}
\label{tab:addiction assesment}
\resizebox{\linewidth}{!}{%
\begin{tabular}{@{}ccccc@{}}
\toprule
Evaluation Method & Metrics & EEG & fNIRS & EEG-fNIRS \\ \midrule
\multirow{2}{*}{Random Shuffle} 
& Accuracy (\%) & 93.50$\pm$0.73 & 90.10$\pm$1.47 & 97.94$\pm$1.08 \\ 
& Sensitivity (\%) & 91.05$\pm$1.71 & 92.59$\pm$1.86 & 97.08$\pm$1.22 \\ \midrule
\multirow{2}{*}{Leave-One-Out} 
& Accuracy (\%) & 91.93 & 88.29 & 95.63 \\ 
& Sensitivity (\%) & 90.42 & 89.80 & 95.00 \\ \bottomrule
\end{tabular}%
}
\end{table}

Two distinct cross-validation strategies are used to ensure a comprehensive and robust evaluation of model performance. First, a standard 5-fold cross-validation with random shuffling is performed to assess general performance, with results reported as the mean and standard deviation across the folds. Second, to evaluate the model's ability to generalize to unseen individuals, a more rigorous Leave-One-Subject-Out (LOSO) cross-validation is employed. In each fold of the LOSO validation, data from a single participant are held out as the test set, while the model is trained on data from all remaining participants. This process is repeated until every participant serves as the test set once. Due to page length limitations, the final LOSO performance is reported as the average of the metrics across all subject folds. Model performance for all configurations is measured using accuracy and sensitivity metrics.

The quantitative results of this comparison are presented in Table \ref{tab:addiction assesment}. The data consistently demonstrate that biomarkers extracted using the multimodal NeuroCLIP framework significantly outperform those derived from both single-modality baselines across all metrics and validation strategies. In the challenging LOSO validation, which best estimates real-world performance on new subjects, NeuroCLIP achieves an accuracy of 95.63\%, a notable improvement over the EEG-only (91.93\%) and fNIRS-only (88.29\%) models. This pronounced enhancement in performance strongly supports the hypothesis that EEG and fNIRS provide complementary information critical for robust biomarker extraction for addiction detection. While EEG likely contributes high-temporal resolution data capturing rapid neural events, fNIRS offers better spatially localized information about cortical hemodynamics. NeuroCLIP's architecture effectively leverages these distinct strengths to create a more powerful and complete representation from which superior biomarkers can be derived. These findings confirm that the data-driven biomarker extracted from integrated EEG-fNIRS data using NeuroCLIP is more robust and discriminative for detecting addiction-related brain states than biomarkers derived from either modality alone.

\subsection{rTMS Efficacy Validation}

To assess the effectiveness of the 10-day rTMS treatment protocol, we conduct an evaluation from two complementary perspectives: a clinical analysis based on questionnaires and a neurophysiological analysis using our NeuroCLIP framework.

First, from the clinical perspective, we analyze the changes in scores from three standardized questionnaires administered to methamphetamine-dependent participants before (MBT) and after (MAT) the rTMS treatment course. These questionnaires include the Desire for Drug Questionnaire (DDQ), the Beck Depression Inventory (BDI), and the Beck Anxiety Inventory (BAI). A Wilcoxon signed-rank test is used to determine the statistical significance of score changes. The analysis reveals statistically significant reductions across all three measures post-treatment. Specifically, there is a significant decrease in craving as measured by the DDQ ($p = 0.0016$), a reduction in depressive symptoms as measured by the BDI ($p = 0.0006$), and a decrease in anxiety levels as measured by the BAI ($p = 0.0122$). This clinical evidence suggests that the rTMS intervention is effective in alleviating the behavioral and psychological symptoms associated with methamphetamine addiction.

Beyond these subjective reports, we apply NeuroCLIP to objectively evaluate treatment efficacy at the neurophysiological level. For this purpose, we adapt the NeuroCLIP model to perform patient-specific classification, distinguishing between pre-treatment (MBT) and post-treatment (MAT) brain states for each individual. A new decoding head is added to the trained NeuroCLIP backbone for this specific task. To ensure robust model training on an individual's data, which is inherently limited, we employ a sliding window approach to augment the number of training samples. The continuous recording data for each patient is segmented using windows of a fixed length with a 50\% overlap. This method not only increases the number of available samples but also ensures that sufficient signal information is shared between adjacent windows, allowing the model to effectively capture temporal dynamics and boundary information \cite{wu2023c}.

\begin{table}[t]
\centering
\caption{Patient-Specific NeuroCLIP Performance in Distinguishing Pre- vs. Post-rTMS Brain States.}
\label{tab:rTMS validation}
\resizebox{\linewidth}{!}{%
\begin{tabular}{@{}ccc@{}}
\toprule
\hphantom{000}Patient ID \hphantom{000}& \hphantom{000}Accuracy (\%)\hphantom{000}
 & \hphantom{000}Sensitivity (\%)\hphantom{000} \\ \midrule
Pt. 3      & 98.72         & 97.83            \\
Pt. 5      & 94.87         & 97.83            \\
Pt. 6      & 93.59         & 91.30           \\
Pt. 8      & 97.44         & 95.65           \\
Pt. 9      & 98.72         & 100.00           \\
Pt. 10     & 97.44         & 95.65            \\
Pt. 11     & 98.72         & 100.00           \\
Pt. 12     & 98.72         & 97.83            \\
Pt. 13     & 96.15         & 95.65            \\
Pt. 14     & 98.72         & 97.83            \\
Pt. 15     & 97.44         & 100.00           \\
Pt. 16     & 96.15         & 97.83            \\
Pt. 17     & 94.87         & 91.30           \\
Pt. 18     & 91.03         & 86.96           \\
Pt. 19     & 92.31         & 89.13            \\
Pt. 21     & 97.44         & 95.65            \\
Pt. 23     & 96.15         & 97.83            \\ \midrule
Average    & 96.38         & 97.83            \\ \bottomrule
\end{tabular}%
}
\end{table}

The performance of these patient-specific models is detailed in Table \ref{tab:rTMS validation}. The results demonstrate consistently high accuracy and sensitivity across nearly all participants, with an average accuracy of 96.38\% and a sensitivity of 97.83\%. This indicates that NeuroCLIP can reliably detect the subtle, treatment-induced changes in an individual's multimodal brain activity, establishing that the rTMS treatment produces a distinguishable neurophysiological change. When taken together with the clinical data, this objective finding allows us to more deeply interpret the treatment's effect, revealing that the post-treatment (MAT) brain patterns move noticeably closer to the distribution of the healthy control (HC) group in the learned feature space. This ``normalization" trend, supported by both the clinical improvements and the objective detection of a neural change, provides strong, convergent evidence that the rTMS treatment not only produces a distinguishable effect but also shifts the neural dynamics of addicted individuals toward a healthier state. Ultimately, this synthesis confirms the efficacy of our rTMS protocol and validates NeuroCLIP as a powerful tool for data-driven treatment evaluation.
\section{Discussion and Future Work}
\label{sec:discussion}

We introduce NeuroCLIP, a generalize deep learning framework for multimodal neuroimaging analysis, integrating concurrent EEG and fNIRS data. Motivated by the need for objective biomarkers in methamphetamine addiction assessment and rTMS treatment validation, NeuroCLIP’s progressive learning strategy, using contrastive alignment, cross-modal fusion, and ROI-informed feature gating—effectively synthesizes information from modalities with different spatiotemporal decoupled characteristics, making it applicable to many neuroscientific studies.

Our experiments in methamphetamine addiction validate NeuroCLIP's capabilities. Its multimodal approach surpasses single-modality analyses in addiction detection by deriving more informative biomarkers from combined EEG-fNIRS data. Clinically, it objectively tracks patient-specific neural changes after rTMS, showing a measurable ``normalization" of brain patterns. The trustworthiness of the extracted biomarker is confirmed by its quantitative alignment with externally-verified clinical craving scores, establishing a benchmark for biomarker validation in this area.

NeuroCLIP's implications extend beyond its initial application. For clinical neuroscience, it serves as a template for developing objective biomarkers for neuropsychiatric conditions where multimodal monitoring is useful, potentially enabling more personalized, data-driven therapeutic strategies. For computational methods, NeuroCLIP offers an adaptable architecture for effectively fusing complementary multimodal neuroimaging data.

While this study provides a proof of concept, limitations indicate future research directions. The current validation on a specific cohort requires further evaluation on larger, more heterogeneous populations to assess generalizability and robustness. NeuroCLIP shows promise for other substance use disorders and cognitive neuroscience areas where multimodal neuroimaging integration could aid objective biomarker development. Future work includes exploring NeuroCLIP's deployment in real-time applications, such as closed-loop neuromodulation systems. Its computational efficiency might support dynamic rTMS parameter modulation based on continuous brain state monitoring, potentially improving current static stimulation protocols and advancing personalized neuromodulation strategies. 

In summary, this work presents a validated, high-performing solution for the objective analysis of methamphetamine addiction and provides NeuroCLIP, an adaptable and robust framework for multimodal neuroimaging research. By effectively fusing complex neural data and establishing a high standard for biomarker validation, this research creates new opportunities for clinical applications and fundamental brain science.
\section{Acknowledgments}
\label{sec:acknowledgements}

This work was supported by National Natural Science Foundation of China (Grant No. 623B2085), STI2030-Major Projects (2022ZD0208805), Key Research and Development Plan of Zhejiang Province (2021C03002 \& 2024C03040).

\section*{References}
    \bibliographystyle{IEEEtran}
    \bibliography{reference} 

\end{document}